\documentclass[prl,twocolumn,showpacs,groupedaddress]{revtex4}
\usepackage{colordvi,epsfig,color,amsmath,amssymb}

\begin{document}

\def\be{\begin{equation}}
\def\ee{\end{equation}}
\def\bee{\begin{eqnarray}}
\def\eee{\end{eqnarray}}
\def\sech{\mbox{sech}}
\def\e{{\rm e}}
\def\d{{\rm d}}
\def\L{{\cal L}}
\def\U{{\cal U}}
\def\M{{\cal M}}
\def\T{{\cal T}}
\def\V{{\cal V}}
\def\R{{\cal R}}
\def\kb{k_{\rm B}}
\def\tw{t_{\rm w}}
\def\ts{t_{\rm s}}
\def\Tc{T_{\rm c}}
\def\gs{\gamma_{\rm s}}
\def\tm{tunneling model }
\def\TM{tunneling model }
\def\tilde{\widetilde}
\def\Deltac{\Delta_{0\rm c}}
\def\Deltamin{\Delta_{0\rm min}}
\def\Emin{E_{\rm min}}
\def\tauc{\tau_{\rm c}}
\def\tauac{\tau_{\rm AC}}
\def\tauw{\tau_{\rm w}}
\def\taumin{\tau_{\rm min}}
\def\taumax{\tau_{\rm max}}
\def\de{\delta\varepsilon / \varepsilon}
\def\pF{{\bf pF}}
\def\pFAC{{\bf pF}_{\rm AC}}
\def\halb{\mbox{$\frac{1}{2}$}}
\def\with{\quad\mbox{with}\quad}
\def\und{\quad\mbox{and}\quad}
\def\za{\sigma_z^{(1)}}
\def\zb{\sigma_z^{(2)}}
\def\ya{\sigma_y^{(1)}}
\def\yb{\sigma_y^{(2)}}
\def\xa{\sigma_x^{(1)}}
\def\xb{\sigma_x^{(2)}}
\def\spur#1{\mbox{Tr}\left\{ #1\right\}}
\def\erwart#1{\left\langle #1 \right\rangle}
\newcommand{\bbbone}{{\mathchoice {\rm 1\mskip -4mu l}{\rm 1\mskip -4mu l}{\rm 1\mskip -4.5mu l}{\rm 1\mskip -5mu l}}}

\title{Landau Zener transitions in a dissipative environment: Numerically exact results}

\author{P. Nalbach and M. Thorwart}
\affiliation{Freiburg Institute for Advanced Studies (FRIAS),
Albert-Ludwigs-Universit\"at Freiburg, Albertstr. 19, 79104 Freiburg, Germany}

\date{\today}

\begin{abstract}

We study Landau-Zener transitions in a dissipative environment by means of the numerically exact quasiadiabatic propagator path-integral.
It allows to cover the full range of the involved parameters. We discover a nonmonotonic dependence of the transition probability on the sweep velocity which is explained in terms of a simple phenomenological model. This feature, not captured by perturbative approaches, results from a nontrivial competition between relaxation and the external sweep.

\end{abstract}


\maketitle

Nonadiabatic transitions at avoided level crossings play an essential role in many dynamical processes throughout physics and chemistry. They have been extensively studied both theoretically and experimentally in, e.g., solid state artificial atoms \cite{LZSi2006,LZBe2008,LZKi2008}, spin flips in nanomagnets \cite{Wernsdorfer}, nanocircuit QED
\cite{QED1,QED2}, the dynamics of chemical reactions \cite{Nitzan}, 
adiabatic quantum computation \cite{AQC}, and in Bose-Einstein condensates in optical lattices \cite{Zenesini2009}. 

In the pure Landau Zener (LZ) problem, the dynamics is restricted to two quantum states coupled by a constant tunneling matrix element $\Delta_0$. A control parameter is swept through the avoided level crossing at a constant velocity $v$. The focus is on the final occupation probability of the two states. This problem was solved by various authors independently \cite{LZLa1932,LZZe1932,LZSt1932,LZMa1932}. However, in any physical realization, a quantum system is influenced by its environment leading to relaxation and phase decoherence during time evolution \cite{WeissBuch}. Any manipulation scheme using LZ transitions should thus include such effects.

Many limiting cases for the dissipative LZ problem are analytically tractable \cite{LZKa1998,LZAo1991,LZWu2006,LZZu2008,Pokrovsky07}. It has been, e.g., shown that for fast sweeps, at low temperatures or for strong system-bath couplings, the bath does not influence the LZ probability \cite{LZKa1998,LZAo1991,LZZu2008,LZWu2006}. At high temperatures but slow sweeps, the two-level system will be driven towards equal occupation \cite{LZAo1991,LZKa1998}. The limit of fast quantum noise has been considered in Ref.\ \cite{Pokrovsky07}. However, despite these important efforts, no exact solution for the full parameter space is available up to date.

In this Letter, we investigate the full parameter range of sweep velocities, temperatures, damping strengths and bath cut-off frequencies by means of the quasiadiabatic propagator path-integral (QUAPI) \cite{QUAPI1,QUAPI2}. It allows to include
nonadiabatic as well as non-Markovian effects yielding numerically exact results. Most importantly, we identify a nonmonotonic dependence of the transition probability on the sweep velocity which can be explained in terms of a simple physical picture involving a competition between relaxation and LZ driving. Our numerical results also allow to address the validity of existing perturbative approaches.

Landau Zener transitions around an avoided level crossing of two quantum states are described by the
Hamiltonian ($\hbar=k_B=1$)
\be
H_{LZ}(t) = \frac{\Delta_0}{2}\sigma_{x} +\frac{vt}{2} \sigma_z \, ,
\ee
with the tunneling matrix element $\Delta_0$ and the energy gap between the diabatic states $vt$, changing linearly
in time with sweep velocity $v$.
Here, $\sigma_{x,z}$ are Pauli matrices and the diabatic states are the eigenstates
($|\downarrow\rangle$ and $|\uparrow\rangle$)
of $\sigma_z$. Asymptotically at times $|t|\gg\Delta_0/v$,
the diabatic states coincide with the momentary eigenstates of $H_{LZ}$.
The LZ problem asks for the probability $P_0$ of the system to end up in the ground state at $t=+\infty$, having started in the ground state at $t=-\infty$ (the corresponding one for the excited state is given as $1-P_0$). Its exact solution dates back to the year 1932 \cite{LZLa1932,LZZe1932,LZSt1932,LZMa1932} and is given by
\be
\hspace*{-1mm}P_0(v,\Delta_0)=|\langle\uparrow(\infty)|\downarrow(-\infty)\rangle|^2 = 1 - \exp\left( -\frac{\pi\Delta_0^2}{2v} \right)
\ee

To include environmental fluctuations on LZ transitions, we couple
$H_{LZ}$ to a harmonic bath \cite{WeissBuch}, yielding
\be H(t) = H_{LZ}(t) -\frac{\sigma_z}{2}\sum_k\lambda_k (b_k+b_k^\dagger) +\halb\sum_k \omega_k b_k^\dagger b_k 
\ee
with the bosonic annihilation/creation operators $b_k/b_k^\dagger$.
The bath influence is captured by the spectral function $J(\omega)=2\alpha\omega \exp(-\omega/\omega_c)$, for which
we choose here for definiteness an Ohmic form
with the cut-off frequency $\omega_c$ and the coupling strength $\alpha$ \cite{WeissBuch}.
The Landau-Zener probability for the dissipative problem $P=\spur{U^{-1}_\infty|\uparrow\rangle\langle\uparrow|U_\infty|\downarrow\rangle\langle\downarrow}$ with the time evolution operator $U_\infty={\cal T} \exp[-i\int_{-\infty}^\infty dt H(t)]$
is now a function not only of $\Delta_0$ and $v$, but also of $\alpha, \omega_c$ and the temperature $T$.
In the following, we use $\omega_c=10\Delta_0$ unless specified otherwise. We determine $P$ by applying the numerically exact 
 and reliable QUAPI technique \cite{QUAPI1,QUAPI2} which is well suited 
for driven problems~\cite{QUAPI2}.
\begin{figure}[t]
\epsfig{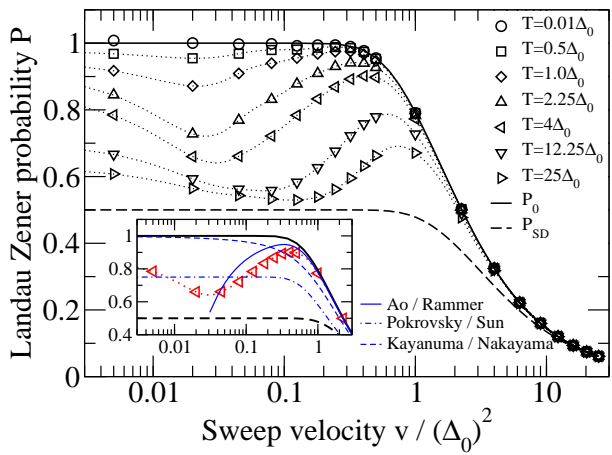}
\vspace*{-1em}\caption{\label{fig1} The LZ probability $P$ for various temperatures $T$ is shown for a weak system-bath coupling $\alpha=0.0016$. The dotted lines are guides to the eye. The solid line marks the coherent LZ probability $P_0$, while the dashed line indicates the high-temperature limit $P_{SD}$, see text. Inset: Comparison with perturbative approaches
for $T=4 \Delta_0$, see text.}
\end{figure}

Figure \ref{fig1} shows the LZ probability $P$ versus sweep velocity $v$ for different temperatures and
for weak coupling, $\alpha=0.0016$. One can distinguish a regime with large velocities, $v\gg \Delta_0^{2}$,
from a regime with small velocities (adiabatic regime) $v\lesssim\Delta_0^{2}$.
Little influence of the bath is expected at low temperatures \cite{LZAo1991,LZKa1998} and it vanishes exactly at $T=0$ \cite{LZWu2006}. This is confirmed by our numerical results.
For small $v$ and low temperatures, $T\lesssim\Delta_0$, we find $P\sim 1$, and thus unmodified compared to the pure quantum mechanical LZ result $P_0$ (solid line). 
For increasing velocity, the LZ probability decreases rapidly and there is hardly any temperature effect in the considered temperature range, see Fig.\ \ref{fig1}. 
This observation agrees with results by Kayanuma and Nakayama who determined the LZ probability in the limit of high temperatures, $P_{SD}=\halb(1-\exp(-\pi\Delta^2/v))$ \cite{LZKa1998} (dashed line),
assuming dominance of phase decoherence over dissipation.
For large velocities, $P_{SD}$ decreases as the pure LZ probability, $P_{SD}\sim P_{LZ}\sim \halb\pi\Delta^2/v$,
and accordingly no sizable temperature effect is expected.

Beyond these limits, in the regime of intermediate to high temperatures, $T>\Delta_0$ and small sweep velocities, $v<\Delta_0^{2}$, we find a nontrivial and unexpected behavior of $P$.
Besides an overall decrease of $P$ with increasing temperature, we find (at fixed $T$) for decreasing velocity first a maximum of $P$ at $v_{\rm max}\lesssim \Delta_0^{2}$, then a minium at $v_{\rm min}$ and finally again an increase (e.g., for $T=4\Delta_0$, $v_{\rm max}\sim 0.5\Delta_0^{2}, v_{\rm min}\sim 0.02\Delta_0^{2}$).
For decreasing temperatures, $v_{\rm min}$ decreases, and $P(v_{\rm min})$ increases.
\begin{figure}[t]
\epsfig{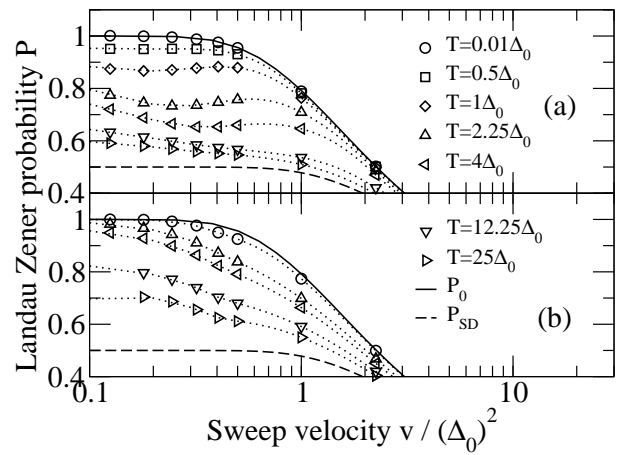}
\vspace*{-1em}\caption{\label{fig2} Same as Fig.\ \ref{fig1} for larger system-bath coupling $\alpha=0.02$ (a) and $\alpha=0.2$ (b).}
\end{figure}

This nonmonotonic feature cannot be described in terms of perturbative approaches.
Ao and Rammer derived temperature-dependent corrections to the LZ probability for low temperatures \cite{LZAo1991}. Their result
is shown in Fig.\ \ref{fig1} (inset, solid blue line). They also found an onset temperature $T_o\propto 1/v$,
above which the temperature affects $P$. Thus, at larger velocities, the decrease of $P$ due to increasing temperature starts at higher temperatures. This is in line with our findings of the maximum in $P(v)$, but it does not account for the minimum and the subsequent increase of $P$ for smaller $v$. In the limit of high temperatures, $P_{SD}\rightarrow 1/2$ for $v<\Delta_0^2$. Thus, $P_{SD}$ captures the decrease of $P$ with increasing temperature, but does not account for the nonmonotonic behavior for decreasing $v$. Similarly, we were not able to match neither Eq.\ (54) of Ref.\ \cite{LZKa1998} nor Eq.\ (40) of Ref.\ \cite{Pokrovsky07} with our  exact data, see Fig.\ \ref{fig1}, inset.

In the following, we provide a simple physical picture of how the bath influences the LZ probability.
We assume that the bath mainly induces relaxation. 
Since initially the system is in the ground-state, only absorption can occur, if an excitation with energy  $\Delta_t=\sqrt{\Delta_0^2+(vt)^2}$ exists in the bath spectrum and is thermally populated. Since $\Delta_t$ is (slowly)
changing with time, relaxation can only occur during a time window $|t|\le\halb t_r$ (with the resonance time $t_r=2\sqrt{\Delta_c^2-\Delta_0^2}/v$),  in which the energy splitting fulfills the condition $\Delta_{t}\le\Delta_c=\mbox{min}\{T,\omega_c\}$ \cite{note1}.
In order for relaxation processes to contribute, the (so far unknown) relaxation time $\tau_r$ must be shorter than $t_r$.

For large sweep velocities, $t_r\hspace*{-0.3mm}\ll\hspace*{-0.3mm}\tau_r$, relaxation is negligible and no influence of the bath is found as expected. In the opposite limit, $t_r\hspace*{-0.3mm}\gg\hspace*{-0.3mm}\tau_r$, relaxation will dominate and the two levels will at any time adjust their occupation to the momentary $\Delta_t$ and $T$. Once $\Delta_t\hspace*{-0.3mm}\ge\hspace*{-0.3mm}\Delta_c$, relaxation stops since no spectral weight of the bath modes is available and the LZ probability is $P_c\hspace*{-0.3mm}=\halb [1+\tanh(\Delta_c/2 T)]$. For small but finite $v$, equilibration is retarded, i.e., equilibrium is reached for an energy splitting in the past $\Delta_{t'}\hspace*{-0.3mm}<\hspace*{-0.3mm}\Delta_t$ with $t'\hspace*{-0.5mm}<\hspace*{-0.5mm}t$ and accordingly, $P$ increases with decreasing $v$ and $P(v\rightarrow0)\hspace*{-0.3mm}\le\hspace*{-0.3mm} P_c$, as observed in Fig.~\ref{fig1}~\cite{note2}.
\begin{figure}[t]
\epsfig{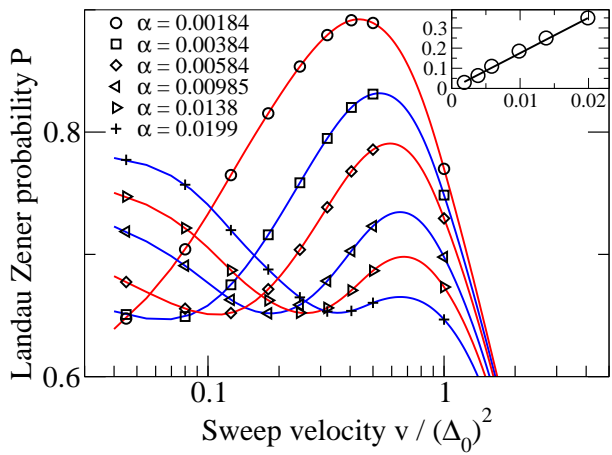}
\vspace*{-1em}\caption{\label{fig3} LZ probability $P$ for different $\alpha$ for $T=4\Delta_0$. Inset: Linear dependence of
$v_{\rm min}$ on $\alpha$.}
\end{figure}

\vspace*{-1em}Relaxation will maximally suppress the LZ transition when both time scales coincide, leading to a minimum of $P$ at $v_{\rm min}$ given by 
\be\label{vmin} t_r(v_{\rm min})=\tau_r(T,\alpha,\omega_c).
\ee
Within resonance, $|t|\le\halb t_r$, only a single phonon absoption is likely. We, thus, assume equillibration towards a time-averaged energy splitting $\overline{\Delta}_r=(2/t_r)\int_0^{t_r/2}dt\Delta_t\simeq\mbox{max}\{\Delta_c/2,\Delta_0\}$
and $P(v_{\rm min})=\halb [1+\tanh(\overline{\Delta}_r/2T)]$. Subsequently, there is a maximum for a sweep velocity between
$v_{\rm min}<v_{\rm max}<\Delta^2_0$.
Fig. \ref{fig5}c shows $P(v_{\rm min})$ vs. $T$ which is in rough agreement with the expected behavior for $T<\Delta_0$ and $T>\omega_c$, but the temperature independent range for $2\Delta_0<T<\omega_c$ is not observed.

For a fixed time and for weak coupling, we can estimate the decay rate out of the ground state using Golden Rule, $\tau^{-1}(t)=\pi\alpha(\Delta_0^2/\Delta_t) \exp(-\Delta_t/\omega_c) n(\Delta_t)$ with the Bose factor $n(\Delta_t)=[\exp(\Delta_t/T)-1]^{-1}$.
For the time-dependent LZ problem at slow sweep velocities, we may
assume that the bath sees a time-averaged two-level system and thus estimate
the relaxation rate $\tau_r^{-1}$ by using the time-averaged energy splitting
$\overline{\Delta}_r$, i.e.,
\vspace*{-0.3em}\be\label{rate} \tau_r^{-1} \,\simeq\, \pi\alpha \frac{\Delta^2_0}{\overline{\Delta}_r}\, \exp(-\overline{\Delta}_r/\omega_c) n(\overline{\Delta}_r)\, .
\ee
For increasing temperature, relaxation becomes faster, and, accordingly, the condition for $v_{\rm min}$, Eq.~(\ref{vmin}), is fulfilled for larger velocities. This describes qualitatively the temperature dependence of the minimum velocity $v_{\rm min}$ in Fig.\ \ref{fig1}. Quantitatively, this simple argument predicts an exponential increase at low temperatures $T\lesssim 2\Delta_0$, a linearly decreasing $v_{\rm min}$ in the intermediate temperature range $2\Delta_0\lesssim T \lesssim \omega_c$, and a linear increase at high temperatures. Fig.\ \ref{fig5}c shows $v_{\rm min}$ versus temperature which nicely fits to the linear behavior $v_{\rm min}=0.019\Delta_0^2+0.004\Delta_0 T$ and thus does not show the expected behavior in the intermediate temperature
range. However, for $T=2\Delta_0$, we estimate $v_{\rm min}=0.028 \Delta_0^2$ and for the linear slope in the high temperature range $0.0024 \Delta_0$, consistent with the data.

\begin{figure}[t]
\epsfig{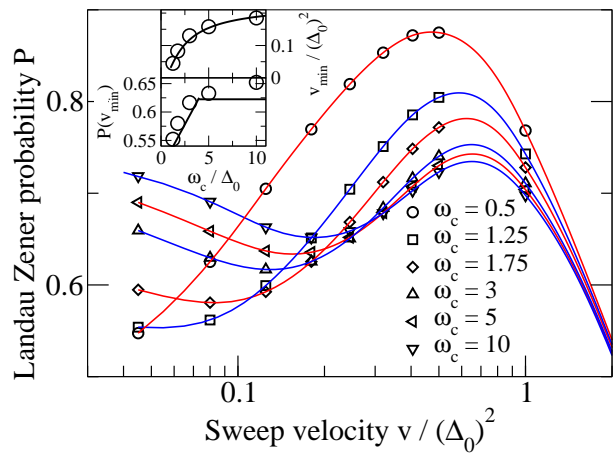}
\vspace*{-1em}\caption{\label{fig4} LZ probability $P$ for various cut-off frequencies $\omega_c$ for $T=4\Delta_0$ and $\alpha=0.00985$. Inset: $P(v_{\rm min})$ (bottom) and $v_{\rm min}$ (top) vs $\omega_c$; solid lines: predictions from our simple model.}
\end{figure}
Clearly, this simple picture breaks down for large sweep velocities when the system changes its energy splitting on a much faster time scale than relaxation time. Thus, a momentary rate and even an averaged rate is not an appropriate description anymore as it requires
that at high temperatures and strong coupling, relaxation should be fast enough to partly equilibrate the system also for fast sweep velocities resulting in $P\rightarrow\halb$. Moreover, we notice that
Kayanuma and Nakayama \cite{LZKa1998} assume dominance of phase decoherence over relaxation in the high temperature limit.
Our findings show that, whereas at very high temperatures or large sweep velocities, phase decoherence is dominant, relaxation dominates the dynamics at intermediate temperatures for small sweep velocities.

Increasing $\alpha$ enhances relaxation and $\tau_r$ decreases. Thus, the minimum sweep velocity is larger for a larger $\alpha$ for a fixed temperature. This is confirmed by Fig. \ref{fig2} where $P$ is shown for the same temperatures as in Fig.\
\ref{fig1}, but for $\alpha=0.02$ (a) and $\alpha=0.2$ (b).
The minimum is still observable for $\alpha=0.02$ for temperatures $\Delta_0\lesssim T\lesssim 4\Delta_0$.
At higher temperatures, only a shoulder remains. For $\alpha=0.2$, the local extrema disappear, but a monotonic growth of $P$ with decreasing sweep velocity is still in line with our simple picture. For coupling strengths $\alpha\ge1/\sqrt{2}$ no bath influence is expected anymore~\cite{LZAo1991}, consistent with our data. Therefore, we focus on $\alpha\le0.2$. 
We only expect a minimum (see Eq.~(\ref{vmin})) for temperatures below the crossover temperature $T_c$ given by $v_{\rm min}(T_c)=\Delta_0^2$. 
$T_c$ is well above the highest investigated temperature for $\alpha=0.0016$ and well below $\Delta_0$ for $\alpha=0.2$. For the intermediate coupling $\alpha=0.02$, $4\Delta_0\lesssim T_c\lesssim 12.25\Delta_0$ and for $\alpha=0.006$ $T_c\simeq 25\Delta_0$ (not shown). $T_c$ increases roughly linearly with growing $\alpha$ as predicted by our model, but the absolute values are off by a factor of three.
 
Surprisingly, our simple picture still holds qualitatively for stronger damping, when the Golden Rule is not expected to hold.
Fig.\ \ref{fig3} shows the LZ probability for different $\alpha$ at a fixed temperature $T=4 \Delta_0$. For increasing
coupling, $v_{\rm min}$ shifts to larger velocities. In fact, $v_{\rm min}$ depends linearly on $\alpha$, see inset of
Fig.\ \ref{fig3}. The linear dependence is also predicted by our model, i.e., $v_{\rm min}=15.8\alpha\Delta_0^2$,
in very good agreement with the fit $v_{\rm min}=17.54\alpha\Delta_0^2$.
The decreasing maximum $P(v_{\rm max})$ results from the shifting minimum.
Another remarkable fact is that the LZ probability $P(v_{\rm min})$ at the minimum velocity is independent of $\alpha$,
as expected. We estimate the averaged splitting $\overline{\Delta}_r=T/2=2\Delta_0$, in fair agreement with $\overline{\Delta}_r=2.476\Delta_0$, obtained with $P(v_{\rm min})=\halb [1+\tanh(\overline{\Delta}_r/2T)]$ from the data in Fig.\ \ref{fig3}. This agreement strongly supports our conclusion that relaxation dominates the LZ probability in the intermediate temperature range for small sweep velocities.

The relaxation rate (\ref{rate}) also depends on the cut-off frequency $\omega_c$ of the bath spectrum and relaxation is strongly suppressed when $\Delta_t>\omega_c$. Fig.\ \ref{fig4} shows $P$ for different $\omega_c$, ranging down to $\omega_c =0.5 \Delta$, a situation occurring for biomolecular exciton dynamics in a protein-solvent environment \cite{BioTh2008}.
With decreasing cut-off frequencies, the minimum in $P(v)$ shifts to smaller $v$ as qualitatively expected from Eq.\ (\ref{rate}) (note that this is surprising since a small $\omega_c$ also induces strong non-Markovian effects).
At the same time the LZ probability $P(v_{\rm min})$ decreases.
With decreasing $\omega_c$, the resonance time $t_r$ and the averaged energy splitting $\overline{\Delta}_r$ decrease. For cut-off frequencies $T\le\omega_c$, we expect $v_{\rm min}=0.19\exp(-2\Delta_0/\omega_c)$, in fair agreement with the fit $v_{\rm min}=0.23\exp(-2\Delta_0/\omega_c)$. This is shown in the upper inset in Fig.\ \ref{fig4}. The lower inset shows $P(v_{\rm min})$ versus $\omega_c$. The solid lines are predictions from our model and are in fair agreement with data.
\begin{figure}[t]
\epsfig{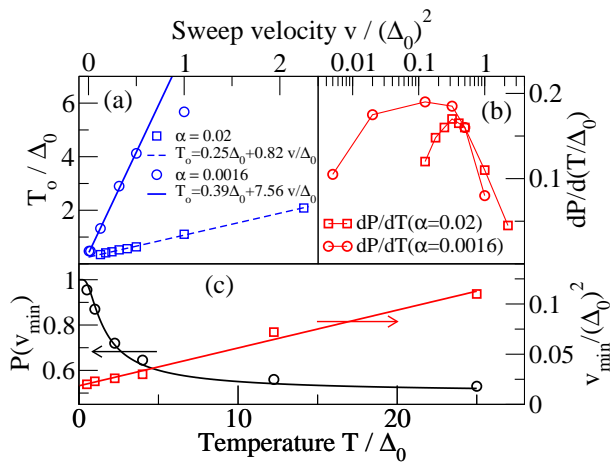}
\vspace*{-1em}\caption{\label{fig5} (a) Onset temperature $T_o$ and (b) maximal slope vs sweep velocity for $\alpha=0.02, 0.0016$.
(c) LZ probability at the minimum sweep velocity (left axis, black circles) and minimum sweep velocity
(right axis, red squares) vs temperature, inferred from Fig.\ \ref{fig1}.}
\end{figure}

Finally, we analyze the dependence of the onset temperature $T_o$ on the sweep velocity $v$. Fig.\ \ref{fig1} suggests that with increasing temperature, $P$ decreases and the onset temperature increases with increasing $v$.
We have extracted $T_o$ from our data by taking the intersection point of a linear fit of $P(T)$ at the point of maximal slope with $P(T=0)=1$. Figure \ref{fig5} (a) shows the onset temperature $T_o$ and Fig.\
 \ref{fig5} (b) the corresponding slope $dP/dT$ versus $v$. The slope $dP/dT$ clearly displays a peak whose height is
only weakly dependent on $\alpha$ and whose width strongly decreases with $\alpha$. The onset temperature increases linearly with $v$ with a constant off-set. This finding is in contrast to the prediction $T_o\sim 1/v$ of Ref.\ \cite{LZAo1991}.

We have investigated the dissipative LZ problem by means of the numerically exact QUAPI \cite{QUAPI1,QUAPI2} approach for an Ohmic bath. In the limits of large and small sweep velocities and low temperatures, our results coincide with analytical predictions \cite{LZAo1991,LZKa1998,LZWu2006,LZZu2008}. For small sweep velocities and medium to high temperatures, however, we have discovered non-monotonic dependencies on the sweep velocity, temperature, coupling strength and cut-off frequency. We have shown that this parameter range is not accessible by perturbative means.
This behavior can be understood in simple physical terms as a nontrivial competition between relaxation and LZ driving. This
novel feature is accessible by available experimental techniques.

We thank V. Peano and S. Ludwig for discussions and acknowledge support by the Excellence Initiative of the German Federal and State Governments.


\begin{thebibliography}{99}
\bibitem{note1} For $\omega_c, T<\Delta_0$, our definition of the resonance time is inept and setting $\Delta_c=\Delta_0$ is more adequate.
\bibitem{note2} For $T=25\Delta_0$ and $\omega_c=10\Delta_0$, we expect $P\lesssim 0.6$, in agreement with Fig.\ \ref{fig1} for the smallest sweep velocity.
\end{thebibliography}
\end{document}